\documentclass[12pt]{iopart}

\usepackage{graphicx}
\usepackage{dcolumn}
\usepackage{bm}

\newcommand{\cm}{cm$^{-1}$}

\newcolumntype{b}{D{(}{\ (}{-1}}

\begin{document}

\title[Sensitivity of deuterated ammonia to
variation of electron-to-proton mass ratio]{Sensitivity of microwave
spectra of deuterated ammonia to variation of electron-to-proton
mass ratio}

\author{M G Kozlov$^1$, A V Lapinov$^2$ and S A Levshakov$^3$}
\address{$^1$Petersburg Nuclear Physics Institute, Gatchina, 188300, Russia}
\eads{mgk@mf1309.spb.edu}
\address{$^2$Institute of Applied Physics, Ul'yanova Str. 46, Nizhny Novgorod, 603950, Russia}
\address{$^3$Ioffe Physical-Technical Institute, Politekhnicheskaya Str. 26, 194021,
St.\ Petersburg, Russia}

\date{\today}

\begin{abstract}

We estimate sensitivity coefficients $Q_\mu$ to variation of the
electron-to-proton mass ratio $\mu$ for microwave transitions in
partly deuterated ammonia NH$_2$D and ND$_2$H. Because of the mixing
between rotational and inversion degrees of freedom the coefficients
$Q_\mu$ strongly depend on the quantum numbers of the transition.
This can be used for astrophysical search for possible variation of
the constant $\mu$.
\end{abstract}

\submitto{\jpb}

\maketitle

\section{Introduction}
\label{intro}

At present discrete microwave spectra of molecules are used for astrophysical
studies of possible variation of the fine structure constant
$\alpha=e^2/(\hbar c)$, the electron-to-proton mass ratio
$\mu=m_\mathrm{e}/m_\mathrm{p}$, and the nuclear $g$-factor $g_\mathrm{n}$.
Rapid progress in experiments with cold molecules can lead to new high
precision laboratory tests of the possible variation of $\mu$ (see, for
example, review \cite{FK09} and references therein). It was pointed out in
\cite{VKB04} that inversion transitions in fully deuterated ammonia
$^{15}$ND$_3$ have high sensitivity to $\mu$-variation, $Q_\mu=5.6$
(sensitivity coefficients $Q$ are defined in \sref{sensitivity}).
The inversion transition in non-deuterated ammonia has a slightly smaller
sensitivity to $\mu$-variation, $Q_\mu=4.5$ \cite{FK07a}. Molecular
rotational lines have much smaller sensitivity, $Q_\mu=1.0$. In astrophysics
the observed frequency shifts are interpreted as Doppler shifts. Because of
that, possible $\mu$-variation would lead to apparent velocity offset between
ammonia inversion line and rotational molecular lines, originated from the
same gas clouds. This fact was used in \cite{FK07a,MFMH08,HMM09} to establish
very stringent limits on $\mu$-variation over cosmological timescale $\sim
10^{10}$ years.

Recently ammonia method was applied to dense prestellar molecular
clouds in the Milky Way \cite{LMK08,MLK09,LML09}. These observations
provide a safe bound of a maximum velocity offset between ammonia
and other molecules at the level of $|\Delta V| \le 28$ m/s. This
bound corresponds to $|\Delta\mu/\mu| \le 3\times 10^{-8}$, which is
two orders of magnitude more sensitive than extragalactic
constraints \cite{FK07a,MFMH08,HMM09}. Taken at face value the
measured $\Delta V$ shows positive shifts between the line centers
of NH$_3$ and other molecules and suggests a real offset, which
would imply a $\Delta\mu/\mu= (2.2\pm 0.4_\mathrm{stat}\pm
0.3_\mathrm{sys})\times 10^{-8}$ \cite{LML09}.
These results can be relevant to the theories, which predict
dependence of the fundamental constants on the local matter density
\cite{BDS09}.

One of the main possible sources of the systematic errors in such
observations is the Doppler noise, i.e. stochastic velocity offsets
between different species caused by different spacial distributions
of molecules in the gas clouds \cite{KCL05,LRK08}. Because of that
it is preferable to use lines with different sensitivity to
variation of fundamental constants of the same species. Recently it
was shown that sensitivity coefficients for $\Lambda$-doublet
spectra of OH and CH molecules strongly depend on quantum numbers
\cite{KC04,Koz09}. In this context partly deuterated ammonia
molecules NH$_2$D and ND$_2$H may be also interesting. Due to the
broken symmetry, the rotational and inversion degrees of freedom for
these molecules are strongly mixed. As we will show below, this
leads to a significant variation of the sensitivity coefficients of
different microwave transitions. Note that microwave spectra of
NH$_2$D and ND$_2$H from the interstellar medium were recently
detected
\cite{OBR85,SOO00,TRF00,RTC00,LCC01,LRT05,GLP06,LGR06,LGR08}.

\section{Sensitivity coefficients}\label{sensitivity}

Let us define dimensionless sensitivity coefficients to the
variation of fundamental constants so that:
 \begin{equation}\label{K-factors}
 \frac{\delta\omega}{\omega}
 = Q_\alpha\frac{\delta\alpha}{\alpha}
 + Q_\mu\frac{\delta\mu}{\mu}
 + Q_g\frac{\delta g_\mathrm{n}}{g_\mathrm{n}}\,.
 \end{equation}

These coefficients $Q_i$ are most relevant in astrophysics, where
lines are Doppler broadened and linewidth
$\Gamma\approx\Gamma_D=\omega\times\delta V/c$, where $\delta V$ is
the velocity distribution width and $c$ is the speed of light. The redshift of a
given line is defined as $z_i=\omega_{\mathrm{lab},i}/\omega_i-1$.
Frequency shift \eref{K-factors} leads to the change in the apparent
redshifts of individual lines. The difference in the redshifts of
two lines is given by:
 \begin{equation}\label{redshifts1}
 \frac{z_i-z_j}{1+z}
 = - \Delta Q_\alpha\frac{\delta\alpha}{\alpha}
 - \Delta Q_\mu\frac{\delta\mu}{\mu}
 - \Delta Q_g\frac{\delta g_\mathrm{n}}{g_\mathrm{n}}\,.
 \end{equation}
where $z$ is the average redshift of both lines and $\Delta
Q_\alpha=Q_{\alpha,i}-Q_{\alpha,j}$, etc. We can rewrite
\eref{redshifts1} in terms of the variation of a single parameter
$\cal F$:
 \begin{equation}\label{redshifts2}
 \frac{z_i-z_j}{1+z}
 = - \frac{\delta{\cal F}}{\cal F}\,,
 \quad
 {\cal F}\equiv
 \alpha^{\Delta Q_\alpha}
 \mu^{\Delta Q_\mu}
 g_\mathrm{n}^{\Delta Q_g}\,.
 \end{equation}
The typical values of $\delta V$ for extragalactic spectra is about
few km/s. This determines the accuracy of the redshift measurements
on the order of $\delta z=10^{-5}$ --~$10^{-6}$, practically
independent on the transition frequency. For gas clouds in the Milky
Way the accuracy can be two orders of magnitude higher, $\delta
z=10^{-7}$ --~$10^{-8}$. In both cases \textit{the sensitivity of
astrophysical spectra to variations of fundamental constants
directly depends on $\Delta Q_i$}.

In the optical range the sensitivity coefficients are typically on
the order of $10^{-2}$ --~$10^{-3}$, while in the microwave and far
infrared frequency regions $Q_i \sim 1$. However, \Eref{redshifts2}
shows, that we need lines with \textit{different} sensitivities. It
is well known that for rotational transitions $Q_\mu = 1.0$, whereas
for vibrational transitions $Q_\mu = 0.5$. For both of them
$|Q_\alpha|\ll 1$ and $|Q_g|\ll 1$. Inversion transition in NH$_3$
has $Q_\mu=4.5$. In the microwave region one can also observe
hyperfine transitions ($Q_\alpha=2$, $Q_\mu=1$, $Q_g=1$) and
$\Lambda$-doublet transitions, where $Q_\alpha$ and $Q_\mu$ strongly
depend on quantum numbers and can be very large \cite{Koz09}. This
makes observations in microwave and far infrared wavelength regions
potentially more sensitive to variations of fundamental constants,
as compared to optical observations. Because of the lower
sensitivity, systematic effects in the optical region are
significantly larger \cite{GWW09}.

\subsection{Inversion transitions}\label{inversion}

Sensitivity coefficients for inversion transitions in NH$_3$ and
ND$_3$ were obtained in \cite{VKB04,FK07a}. For non-symmetric molecules
NH$_2$D and ND$_2$H selection rules are such that purely inversion
transitions are not observable. Still, we will first estimate
sensitivity coefficients for the inversion transition and then will
proceed to the mixed inversion-rotation transitions.

In the WKB approximation the inversion transition frequency is given by
the expression \cite{LL77}:
\begin{equation}
\label{WKB} \omega_\mathrm{inv}=\frac{\omega_\mathrm{v}}{\pi}\rme^{-S}\,,
\end{equation}
where $\omega_\mathrm{v}$ is the vibrational frequency for the inversion mode
and $S$ is the action (we use atomic units $\hbar=m_e=e=1$). Vibrational
frequency is inversely proportional to the square root of the reduced mass
$M_1$ and, hence, is proportional to $\mu^{1/2}$. The action,
 \begin{equation}\label{action}
 S=\int_{-a}^a \sqrt{2M_1[U(x)-E]}\rmd x\,,
 \end{equation}
to a first approximation, is proportional to $\mu^{-1/2}$
(integration here goes between classical turning points, i.e. $U(\pm
a)=E$). Therefore, we can rewrite \eref{WKB} in the form
\cite{VKB04}:
\begin{equation}
 \label{WKB1}
 \omega_\mathrm{inv}=a_1\mu^{1/2}
 \exp\left(-a_2\mu^{-1/2}\right)\,,
\end{equation}
and find the respective sensitivity coefficient:
 \begin{equation}
 \label{K_beta_inv}
 Q_\mu=\case{1}{2}\left(1+a_2\mu^{-1/2}\right)
 =\case{1}{2}\left(1+S\right)\,.
 \end{equation}

Deriving \eref{K_beta_inv} we neglected that the energy $E$ in \eref{action}
also depends on $\mu$, $E=U_\mathrm{min}+\case{1}{2} \omega_\mathrm{v}$.
Taking this into account we get corrected expression \cite{FK07a}:
 \begin{equation}
 \label{K_beta_inv1}
 Q_\mu=\frac{1}{2}\left(1+S
 +\frac{S}{2}\frac{\omega_\mathrm{v}}{\Delta U
 -\case12\omega_\mathrm{v}}\right)\,.
 \end{equation}

Following \cite{SI62} we can estimate
 $\Delta U\equiv U_\mathrm{max}-U_\mathrm{min}$
for ammonia to be approximately 2020 \cm.
Now we can use experimental frequencies $\omega_\mathrm{v}$ and
$\omega_\mathrm{inv}$ for different isotopic variants of ammonia to find $S$
from \eref{WKB} and estimate $Q_\mu$ using \eref{K_beta_inv1}. Results are
presented in \tref{tab1}.

\begin{table}
 \caption{Sensitivity coefficients $Q_\mu$ for the inversion transitions in
 different isotopologues of ammonia.}
 \label{tab1}
 \begin{indented}
 \item[]
 \begin{tabular}{@{}lcccc}
 \br
 \multicolumn{1}{c}{Molecule}
 &\multicolumn{1}{c}{Action}
 &\multicolumn{3}{c}{$Q_\mu$}\\
                 &    $S$  & this work
                                   & \cite{VKB04}
                                           & \cite{FK07a}\\
 \mr
 $^{14}$NH$_3$   &   5.9   &  4.4  &       &  4.5  \\
 $^{15}$NH$_3$   &   6.0   &  4.4  &       &       \\
 $^{14}$NH$_2$D  &   6.5   &  4.7  &       &       \\
 $^{14}$ND$_2$H  &   7.3   &  5.1  &       &       \\
 $^{14}$ND$_3$   &   8.4   &  5.7  &       &  5.7  \\
 $^{15}$ND$_3$   &   8.5   &  5.7  &  5.6  &       \\
 \br
 \end{tabular}
 \end{indented}
\end{table}

\subsection{Mixed transitions}\label{sec_mixed}

For partly deuterated ammonia inversion levels have different
ortho-para symmetry. Because of that inversion transitions can be
observed only in combination with rotational transitions
$\omega_\mathrm{r}$. For such a mixed transition,
 \begin{equation}\label{mixed}
 \omega = \omega_\mathrm{r} \pm \omega_\mathrm{inv}\,,
 \end{equation}
and sensitivity coefficient is equal to:
 \begin{equation}\label{K_mixed}
 Q_\mu =
 \frac{\omega_\mathrm{r}}{\omega} Q_{\mathrm{r},\mu}
 \pm\frac{\omega_\mathrm{inv}}{\omega} Q_{\mathrm{inv},\mu}\,,
 \end{equation}
where $Q_{\mathrm{r},\mu}=1$ and $Q_{\mathrm{inv},\mu}$ is given in
\tref{tab1}.

\Eref{K_mixed} shows that the sensitivity coefficient of the mixed transition
is simply a weighted average of those of constituents. For ammonia all
observed transitions have complex hyperfine structure. That means that in
addition to two term in \eref{mixed} there is third hyperfine term
$\omega_\mathrm{hf}$. However, the hyperfine contribution is typically very
small, $|\omega_\mathrm{hf}/\omega| \ll 1$. Therefore, we can neglect
hyperfine corrections to the sensitivity coefficients. Exceptions to this rule
take place when there are accidental degeneracies between different levels and
respective transition frequencies become comparable to hyperfine structure.
This happens, for example, for some $\Lambda$-doublet transitions in NO and
LiO molecules \cite{Koz09}.

\section{Molecules {NH$_2$D} and {ND$_2$H}}\label{asymmetric_molecules}

Partly deuterated ammonia NH$_2$D and ND$_2$H are asymmetric tops.
Their rotational levels are classified by the rotational quantum
number $J$ and two projection $K_a$ and $K_c$, which correspond to
two limiting cases of prolate and oblate symmetric tops. Thus,
rotational levels are labeled as $J_{K_a,K_c}$. Symmetric and
asymmetric levels, which correspond to the inversion tunneling mode
are labeled with indexes $s$ and $a$ respectively.

Total inversion symmetry of the rotation-inversion levels is
$(-1)^{K_c}$ for $s$ levels and $(-1)^{K_c+1}$ for $a$ levels.
Inversion accompanied by rotation over $\pi$ around one of the
principle axes corresponds to transposition of the two identical
nuclei (H for NH$_2$D, or D for ND$_2$H). The allowed levels are
determined by the total spin of the identical nuclei. Because of
that a purely inversion transition is strongly suppressed as an
\textit{ortho} $\leftrightarrow$ \textit{para} one.

These selection rules lead to two types of allowed transitions. For
transitions with $\Delta K_c=0$ the allowed combination is $s
\leftrightarrow a$ and for transitions with $\Delta K_c=\pm 1$ the
allowed combinations are $s \leftrightarrow s$ and $a
\leftrightarrow a$. In other words, transitions with $\Delta K_c\neq
0$ are pure rotational, while transitions with $\Delta K_c=0$ are
mixed. Consequently, all transitions with $\Delta K_c\neq 0$ have
$Q_\mu=1$, while transitions with $\Delta K_c=0$ according to
\eref{K_mixed} have sensitivities:
 \begin{equation}\label{K_vertical}
 Q_\mu (\omega)= 1 \pm \frac{\omega_\mathrm{inv}}{\omega}
 \left(Q_\mathrm{inv,\mu}-1\right)\,.
 \end{equation}

For NH$_2$D molecule the inversion frequency is 12.2 GHz and
$Q_\mathrm{inv,\mu}=4.7$, so
 \begin{equation}\label{K_vertical1}
 Q_\mu (\omega)= 1 \pm \frac{45\mathrm{GHz}}{\omega}\,.
 \end{equation}
For ND$_2$H molecule the inversion frequency is $\omega_\mathrm{inv}=5.11$ GHz
and $Q_\mathrm{inv,\mu}=5.1$. Thus,
 \begin{equation}\label{K_vertical2}
 Q_\mu (\omega)= 1 \pm \frac{21\mathrm{GHz}}{\omega}\,.
 \end{equation}
Results for several strong transitions between lowest rotational
levels of these molecules are given in tables \ref{tabNH2D}
and~\ref{tabND2H}.

\begin{table}
\caption{Sensitivity coefficients $Q_\mu$ for several lower
$\Delta K_c=0$ microwave transitions in NH$_2$D. The lower state
energy $E$ is given in the third column.}
 \label{tabNH2D}
 \begin{indented}
 \item[]
 \begin{tabular}{@{}lcccc}
 \br
 \multicolumn{2}{c}{transition}
 &\multicolumn{1}{c}{$E$ (\cm)}
 &\multicolumn{1}{c}{$\nu$ (GHz)}
 &\multicolumn{1}{c}{$Q_\mu$}\\
 \mr
 $1_{1,0}(a)\rightarrow 0_{0,0}(s)$  & {\it para}  &   0.00 &  494.5  &  1.09  \\
 $1_{1,0}(s)\rightarrow 0_{0,0}(a)$  & {\it ortho} &   0.41 &  470.3  &  0.90  \\
 $1_{1,1}(a)\rightarrow 1_{0,1}(s)$  & {\it para}  &  11.10 &  110.2  &  1.41  \\
 $1_{1,1}(s)\rightarrow 1_{0,1}(a)$  & {\it ortho} &  11.51 &   85.9  &  0.48  \\
 $2_{1,2}(a)\rightarrow 2_{0,2}(s)$  & {\it para}  &  32.78 &   74.2  &  1.61  \\
 $2_{1,2}(s)\rightarrow 2_{0,2}(a)$  & {\it ortho} &  33.19 &   50.0  &  0.10  \\
 $2_{2,1}(a)\rightarrow 2_{1,1}(s)$  & {\it ortho} &  40.01 &  305.7  &  1.15  \\
 $2_{2,1}(s)\rightarrow 2_{1,1}(a)$  & {\it para}  &  40.41 &  282.1  &  0.84  \\
 \br
 \end{tabular}
 \end{indented}
\end{table}

\begin{table}
\caption{Sensitivity coefficients $Q_\mu$ for $\Delta K_c=0$
transitions in ND$_2$H.}
 \label{tabND2H}
 \begin{indented}
 \item[]
 \begin{tabular}{@{}lcccc}
 \br
 \multicolumn{2}{c}{transition}
 &\multicolumn{1}{c}{$E$ (\cm)}
 &\multicolumn{1}{c}{$\nu$ (GHz)}
 &\multicolumn{1}{c}{$Q_\mu$}\\
 \mr
 $1_{1,0}(a)\rightarrow 0_{0,0}(s)$  & {\it ortho} &   0.00 &  388.7  &  1.05  \\
 $1_{1,0}(s)\rightarrow 0_{0,0}(a)$  & {\it para}  &   0.17 &  378.5  &  0.94  \\
 $1_{1,1}(a)\rightarrow 1_{0,1}(s)$  & {\it para}  &   9.10 &   67.8  &  1.31  \\
 $1_{1,1}(s)\rightarrow 1_{0,1}(a)$  & {\it ortho} &   9.27 &   57.7  &  0.64  \\
 $2_{1,2}(a)\rightarrow 2_{0,2}(s)$  & {\it ortho} &  26.66 &   38.7  &  1.54  \\
 $2_{1,2}(s)\rightarrow 2_{0,2}(a)$  & {\it para}  &  26.83 &   28.6  &  0.27  \\
 $2_{2,1}(a)\rightarrow 2_{1,1}(s)$  & {\it ortho} &  32.59 &  193.0  &  1.11  \\
 $2_{2,1}(s)\rightarrow 2_{1,1}(a)$  & {\it para}  &  32.76 &  183.2  &  0.89  \\
 \br
 \end{tabular}
 \end{indented}
\end{table}

We see that differences between coefficients $Q_\mu$ for ND$_2$H are
somewhat smaller, than for NH$_2$D. In combination with
significantly smaller abundance, this makes ND$_2$H less attractive
candidate for the search of $\mu$-variation. \Tref{tabNH2D} shows
that coefficients $Q_\mu$ for $\Delta K_c=0$ transitions in NH$_2$D
molecule with $J\le 2$ vary from 0.1 to 1.6. Thus, the largest
$Q_\mu$ for deuterated ammonia is almost three times smaller than
$Q_\mu$ for the inversion line in NH$_3$. However, in NH$_3$ there
are no close lines with different sensitivities and we are forced to
use rotational lines of other molecules as a reference. For
deuterated ammonia there are four pairs of lines for ortho and para
molecules which have close frequencies and excitation temperatures,
but significantly different sensitivities. For two of these pairs
$\Delta Q_\mu \sim 1$. In addition, there are several transitions
with $\Delta K_c=\pm 1$, which all have $Q_\mu=1$. This gives very
characteristic pattern of frequency variation caused by variation of
$\mu$ and can be used to suppress various systematic effects.
Consequently, reliability of the results can be significantly
improved. We conclude that microwave spectra of deuterated ammonia
can be used as important supplement to the ammonia method used
previously in \cite{FK07a,MFMH08,HMM09,LMK08,MLK09,LML09}.

\ack

This research is partly supported by RFBR grants 08-02-00460,
09-02-00352, and 09-02-12223.

\section*{References}

%
\providecommand{\newblock}{}

\end{document}